\documentstyle[11pt]{article}

\input epsf 
\begin{document}
\pagestyle{empty}
\title{\Large \bf Approximate Treatment of Lepton Distortion in 
Charged-Current Neutrino Scattering from Nuclei}
\author{ \\ Jonathan Engel \\ 
{\small Department of Physics and Astronomy, University of North 
Carolina, Chapel Hill, NC 27599-3255}}
\maketitle
\begin{abstract}
{\normalsize
The partial-wave expansion used to treat the distortion of scattered electrons
by the nuclear Coulomb field is simpler and considerably less time-consuming
when applied to the production of muons and electrons by low and
intermediate-energy neutrinos.  For angle-integrated cross sections, however,
a modification of the ``effective-momentum-transfer" approximation seems to
work so well that for muons the full distorted-wave treatment is 
usually unnecessary,
even at kinetic energies as low as an MeV and in nuclei as heavy as lead.
The method does not work as well for electron production at low energies, but
there a Fermi function usually often proves perfectly adequate.  Scattering of
electron-neutrinos from muon decay on iodine and of atmospheric neutrinos
on iron are discussed in light of these results.}
\end{abstract} 

\newpage

\section{Introduction}
\indent

Converting neutrinos with energies below a few hundred MeV into
electrons or muons by scattering them from nuclei is useful in the search for
new physics.  Radiochemical detectors\cite{r:solar} rely on neutrino-nucleus
scattering to measure the apparently too low solar neutrino flux.
Proton-decay experiments, doubling as neutrino observatories, use neutrino
scattering from oxygen or iron to look for oscillations of neutrinos produced
in the atmosphere\cite{r:atmonu1,r:atmonu2}.  And experiments at Los 
Alamos\cite{r:LSND} measure neutrino-carbon cross sections, in part to check 
the flux in a beam that has reportedly produced neutrino oscillations in the 
lab.

Because neutrinos interact so weakly the Born approximation to
neutrino-nucleus scattering should be accurate.  The only difficulty with this
first-order plane-wave approach is that the electrons or muons produced when
the neutrinos interact feel an electrostatic force from the nucleus as they
leave.  In heavy nuclei the Coulomb force can have large effects and even in
light nuclei it changes cross sections noticeably.  The best way to include
the Coulomb interaction is through the distorted-wave Born approximation
(DWBA)\cite{r:DWBA}, which entails an expansion of the outgoing wave function
in partial waves.  In processes similar to neutrino scattering but more often
measured, two approximations help avoid the cumbersome and time-consuming
partial-wave expansion.  Calculations of beta decay, which produces low-energy
electrons, contain a Fermi function that multiplies the decay amplitude by the
ratio of Coulomb to free Dirac s-wave functions at the nuclear
radius\cite{r:Behrens}.  In electron scattering, at energies much larger than
the electrostatic potential energy in the nucleus, an ``effective momentum
approximation" (EMA) uses plane waves with shortened wavelengths and increased
amplitudes in place of incoming and outgoing distorted waves\cite{r:Pacati}.
Both approximations are substantial improvements over the unrenormalized
plane-wave impulse approximation, though the EMA predictions for differential
cross sections still differ significantly from those of the full DWBA in heavy
nuclei, even at high energies\cite{r:Wright}.

This paper discusses the application of these very convenient approximations
and modifications thereof to charged-current neutrino-nucleus scattering, for
which the outgoing lepton is either an electron or muon.  Because the neutrino
is not charged and the weak interaction is short-ranged, the expansion in
partial waves is less complicated and time-consuming than in electron
scattering.  But thus far the most significant observable in all the neutrino
experiments cited above is the total cross section, and one might question
whether the full DWBA is really required to obtain it.  The two approximations
just discussed preserve the intuitive plane-wave picture of the scattering and
are so much easier to apply that one really ought to retain
them if possible.  On the other hand, it is not clear that they will always
work well; the EMA, for example, has been derived only for extremely
relativistic particles\cite{r:Yennie,r:Lenz,r:Knoll}, while the most energetic
outgoing muons considered here are far from the relativistic limit.  I show
below however that although the Fermi function is accurate only for low-energy
electrons in most nuclei, a simple modification allows the EMA to work
remarkably well (better than in electron scattering), even in heavy nuclei
and, for muons, even at very low energies.

\section{General Considerations}
\indent

The differential cross section for a neutrino of momentum $\vec{k}_{\nu}$ to 
produce a muon or electron at scattered angle $\Omega$ and energy $E$, with 
the nucleus going from state $|i\rangle$ to state $|f\rangle$, can be written 
as a phase space factor times a sum of squared matrix elements, viz:
\begin{eqnarray}
{{\partial^2 \sigma } \over {\partial E \partial \Omega}} = {{G^2  kE} \over
{4 \pi^2}} \sum_{s} | {\cal M} (i \rightarrow f) |^2 \\
{\cal M}_{s} (i \rightarrow f) = \int d^3r ~ \bar{\psi}_{\vec{k},s}^-(\vec{r})  
\gamma_{\mu} \nu_{\vec{k}_{\nu}} e^{i\vec{k_{\nu}} \cdot \vec{r}} 
\langle f|{\cal J}^{\mu}(\vec{r})|i \rangle ~~,
\label{e:1}
\end{eqnarray}
where $G$ is the Fermi coupling constant, ${\cal J}$ is the charged weak 
nuclear current, $\nu_{\vec{k}_{\nu}}$ is the 4-spinor containing 
the momentum-space components of the left-handed neutrino plane wave with
momentum $\vec{k}_{\nu}$, and $\psi_{\vec{k},s}^-(\vec{r})$ is the charged 
lepton 
(electron or muon) wave function with spin projection $s$ and energy $E = 
\sqrt{k^2 + M^2}$ in the electrostatic potential
\begin{eqnarray}
V = & \frac{-Z\alpha}{2R} \left( 3 - r^2/R^2\right) & ~~r < R \nonumber \\
V = & -Z\alpha/r                                      &~~ r > R~~. 
\label{e:v}
\end{eqnarray}
($R$ is the nuclear radius.)  The wave function $\psi^-$ is an eigenstate of 
the full Hamiltonian, electrostatic potential included, that looks 
asymptotically like a plane plus an ingoing spherical wave.  Expanded in 
partial waves, it has the form\cite{r:Akh,r:Dirac}
\begin{equation}
\psi_{\vec{k},s}^- = \frac{1}{\sqrt{\pi} k} \sqrt{\frac{E+M}{E}}
\sum_{l,j,m} i^l \langle l m-s, \frac{1}{2} s | j m \rangle Y_{lm}^*(\hat{r}) 
e^{-i \delta_{j,l}} \frac{1}{r} \left( \begin{array}{c}
                                     iP_{j,l}^{E}(r) \Omega_{l,jm}(\hat{r}) \\
                                      Q_{j,l}^{E}(r) \Omega_{2j-l,jm}(\hat{r})
                                      \end{array} \right) ~~,
\end{equation}
where the
quantum number $j$ and the label $l$ determine\footnote{The total angular
momentum $j$ is a good quantum number but the orbital angular momentum $l$ is 
not and serves as
a label here.} the the eigenvalue $\kappa=(l-j)(2j+1)$ of the operator $K
=-\beta (2\vec{L} \cdot \vec{S}+1)$ ($\beta$ is the usual Dirac-equation
matrix); $\Omega_{l,jm}(\hat{r}) \equiv [Y_l \chi]^j_m$ is a two-spinor with
non-relativistic quantum numbers $l$, $s=\frac{1}{2}$, and $j$; and
$\delta_{j,l}$ is the ``inner phase shift"\cite{r:Dirac}.  The radial wave 
functions
$P_{j,l}^{E}(r)$ and $Q_{j,l}^{E}(r)$ obey the equations
\begin{eqnarray}
\frac{dP_{j,l}^E(r)}{dr} & = &  -\frac{\kappa}{r} P_{j,l}^E(r) - 
[E+M-V(r)]Q_{j,l}^{E}(r) \nonumber \\
\frac{dQ_{j,l}^E(r)}{dr} & = &  \frac{\kappa}{r} Q_{j,l}^E(r) - 
[E-M-V(r)]P_{j,l}^{E}(r) ~.
\label{e:radial}
\end{eqnarray}

Since the goal here is to test approximations that
preserve the easily interpreted plane-wave formalism, I will use
only the vector charge and not the full current in Eq.\ (\ref{e:1}); the 
charge, for which
full distorted-wave cross-section formulae can be displayed and calculated
simply, should be sufficient.  If the nuclear target states have SU(4)
symmetry the major part of the axial-vector current will affect the matrix
element in essentially the same way as the charge.  There is no obvious reason
to think that the rest of the current will change the results in a
qualitative way.  

Using the upper and lower radial wave functions $P_{jl}^E(r)$ and 
$Q_{jl}^E(r)$ from Eq.\ (\ref{e:radial}), one can define
\begin{eqnarray}
F_{LJ,j}^{\pm} & = \int d^3 r ~\frac{1}{r} P_{j,j\pm\frac{1}{2}}^E(r) j_L 
(k_{\nu}r) Y_{J0}(\hat{r}) \rho_{fi}(\vec{r}) \nonumber \\
G_{LJ,j}^{\pm} & = \int d^3 r ~\frac{1}{r} Q_{j,j\pm\frac{1}{2}}^E(r) j_L 
(k_{\nu}r) Y_{J0}(\hat{r}) \rho_{fi}(\vec{r}) ~~,
\label{e:0+}
\end{eqnarray}
where 
\begin{equation}
\rho_{fi}(\vec{r}) = \langle f | \sum_n \delta (\vec{r}-\vec{r_n}) \tau^+_n 
|i\rangle
\label{e:rho}
\end{equation}
is the isovector transition transition density and the sum is over nucleons 
$n$. The angle-integrated cross section, which is of particular interest, 
takes a relatively simple form if the recoil of the nucleus is neglected, 
since $k$ and $E$ are then held 
fixed while $\Omega$ is integrated over, with the result that the cross 
section does not depend directly on the phase shifts and the partial waves do 
not interfere with one another.  The general expression for a final state with 
arbitrary $J^{\pi}$ (and initial state with $0^+$), restricting the 
current to the vector charge, is
\begin{eqnarray}
\frac{d\sigma}{dE} =  {G^2} \frac{E+m}{2k} 
\times 
 ~~~~~~~~~~~~~~~~~~~~~~~~~~~~~~~~~~~~~~~~~~~~~~~~~~~~&&\\
\sum_{j,L}
\left( \begin{array}{l}
       \hat{j}^2 \hat{L}^2
                       \left[ 
                       \langle j-\frac{1}{2} 0,L0|J0 \rangle^2 
                       ({F_{LJ,j}^{-}}^2 +{G_{LJ,j}^{+}}^2) 
                       +\langle j+\frac{1}{2} 0,L0|J0 \rangle^2
                       ({F_{LJ,j}^{+}}^2 +{G_{LJ,j}^{-}}^2) \right]  \\
                       \\
     -2 \langle j+\frac{1}{2} 0,L0|J0 \rangle 
                 \left[ \begin{array}{l}
               \sqrt{[J^2-(L+\frac{1}{2}-j)^2][(J+1)^2-(L+\frac{1}{2}-j)^2]}\\
                 \langle j-\frac{1}{2} 0,L+1~0|J0 \rangle 
                 (F_{L+1J,j}^{-} G_{LJ,j}^{-} -F_{LJ,j}^{+} G_{L+1J,j}^{+})\\ 
                \\
             +  
\sqrt{[J^2-(L+\frac{1}{2}+j)^2][(J+1)^2-(L+\frac{1}{2}+j)^2]}\\
                 \langle j-\frac{1}{2} 0,L-1~0|J0 \rangle 
                 (F_{L-1J,j}^{-} G_{LJ,j}^{-} -F_{LJ,j}^{+} G_{L-1J,j}^{+} )
                 \end{array}  \right] 
         \end{array} \right)~~. \nonumber
\label{e:xsec}
\end{eqnarray}
This expression will be used below for comparing exact results with 
approximations.  I will take the density in Eq.\ (\ref{e:rho}) to have the 
form $\rho_{fi} \propto \delta(r-R) Y_{JM}(\hat{r})$ for transitions from a 
$0^+$ ground state to excited states with angular momentum $J$ and projection 
$M$, so that, e.g., $1^-$ states correspond to the isobar analogs of 
Goldhaber-Teller collective-dipole excitations.
There are several reasons for this choice: densities for most low-lying 
collective states are surface-peaked, placing the strength at the nuclear 
radius provides a 
severe test for the approximations below (which work best at small $r$) and, 
finally, densities for higher-energy noncollective states can
be represented as superposed densities peaked at different points in the 
nucleus, so that a $\delta$-function density should be sufficiently general.
\section{The Fermi Function}
\indent

It is convenient to think about the effects of an electrostatic potential on 
an outgoing particle in terms of a local effective energy and momentum inside 
the nucleus: 
\begin{equation}
E_{\rm eff} = E - V(0), ~~~~~~~~k_{\rm eff} = \sqrt{E_{\rm eff}^2 - M^2}
\end{equation}
When $k_{\rm eff} R << 1$, a Fermi function of $Z$ and $E$ that multiplies the 
outgoing wave 
is appropriate.  In this low ``effective-momentum" limit, only s-waves 
contribute to the scattering, and they vary nearly linearly inside the 
nucleus; the Fermi function can then be taken to be the ratio of, e.g., the 
Coulomb wave to the free wave at the nuclear surface, a quantity given by
\begin{equation}
F(Z,E) = 2(1+\gamma_0)(2kR)^{-2(1-\gamma_0)} e^{\pi\nu} 
\frac{|\Gamma(\gamma_0+i\nu)|^2}{\Gamma(2\gamma_0+1)^2}~~,~~\gamma_0=
\sqrt{1-Z^2\alpha^2} ~~,~~\nu=\pm \frac{Z \alpha E}{k}~.
\label{e:fermi}
\end{equation}  

\begin{figure}[thb]
\caption{\protect The total cross section for 
scattering from the ground states of $^{12}$C and $^{208}$Pb to (fictitious) 
$0^+$ states at 15 MeV in $^{12}$N and 
$^{208}$Bi, with transition densities proportional to $\delta(r-R)$, as a 
function of outgoing lepton kinetic energy for both electrons (a) and muons 
(b).  The solid lines represent the full DWBA results, the dotted lines 
neglect the Coulomb force completely, and the dashed lines are the 
Fermi-function approximations.}
\vspace{3mm}
\centering
\leavevmode
\epsfxsize=10cm
\epsfbox{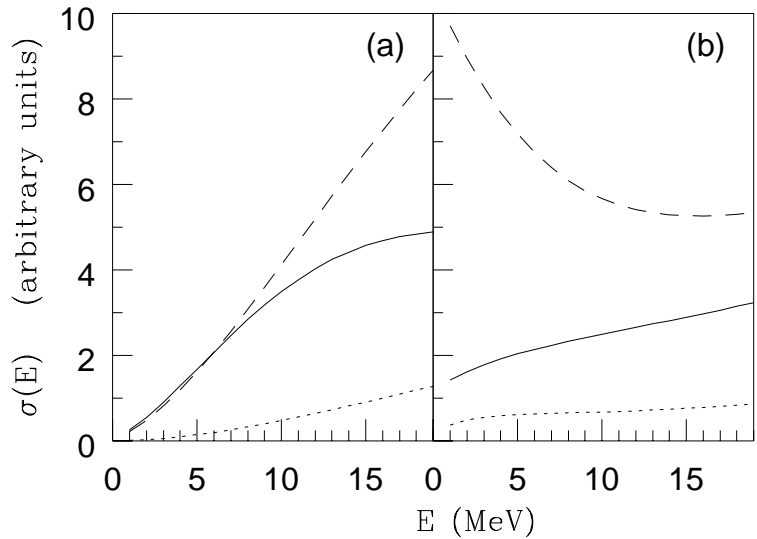}
\label{f:efermi}
\end{figure}

The Fermi function is useful for low-energy electrons and is often employed 
 in calculations of beta decay, but is not likely to work well for muons
except in light nuclei.  The reason is that for nonrelativistic muons, $k_{\rm
eff} \approx \sqrt{2M_{\mu}E_{\rm eff}} \geq \sqrt{2M_{\mu}V_0}$, which even 
in
$^{12}$N, the product of a charge-exchange reaction on $^{12}$C, implies that
$k_{\rm eff} R ~\raisebox{-.25ex}{$\stackrel{<}{\scriptstyle \sim}$}\ .5$
(large enough to cause a 10\% error with the Fermi function above).  Thus,
even when the energy at infinity is so low that only s-wave muons are produced
the usual Fermi function will generally not be not very accurate.  Figure
\ref{f:efermi} shows the total cross section for exciting a (fictitious) $0^+$
state in $^{208}$Bi at 15 MeV with a surface-peaked transition density, as a
function of outgoing lepton kinetic energy for both electrons and muons; the
$0^+$ multipole is the one with the largest s-wave contribution at low
energies so the choice should maximize the accuracy of the Fermi function.
Indeed, for electrons, the function does well for the lowest 5 -- 10 MeV of
electron energy.  For muons, however, the approximation is never valid.  Its
accuracy can be improved significantly by using the s-wave solution from the
potential in Eq.\ (\ref{e:v}), rather than the pure Coulomb potential (see
Ref.\ \cite{r:Behrens}), but except at low energies where such functions are
tabulated, that means solving the s-wave Dirac equation, a task that is 
simpler than carrying out the full DWBA but still relatively involved.

One physical situation in which these issues are important is the
scattering of electron neutrinos produced by muon-decay.  A few years ago an
attempt was made to calibrate an $^{127}$I solar-neutrino detector by exposing
it to a flux of muon-decay neutrinos at LAMPF\cite{r:Lande}.  The neutrinos 
have
much higher energy than their solar counterparts and excite the nucleus in
different ways, so that a good understanding of low-lying states in $^{127}$Xe
is required to extract the solar-neutrino cross section.  In addition, the
outgoing electron often has too high an energy to allow use of the usual Fermi
function (but not high enough for the EMA); if the transition density is
surface-peaked and concentrated at an average of 6 MeV, the Fermi function
gives values about 50\% larger than the DWBA $0^+$ multipole cross section 
when only the vector charge is included.  The difficulties
with nuclear structure and electron distortion were discussed in ref.\
\cite{r:iodine}; to treat the distortion for the $0^+$ and $1^+$ components of 
the cross section, the authors relied on a fit\cite{r:Vogel} to tables in 
Ref.\ \cite{r:Behrens}, which use the potential in Eq.\ (\ref{e:v}) rather 
than the pure Coulomb force to calculate the Fermi function.  Here that 
procedure still misses the cross sections, though only by about 10 or 15\%.  
The difference is due mainly to higher partial waves, which are affected
differently by the electrostatic force than the s-wave that underlies the
Fermi function.  It would take a significantly larger difference, however, to
alter the conclusion of Ref.\ \cite{r:iodine} that the LAMPF cross sections
are too large to be understood or interpreted.

\section{The EMA and a Modified EMA}
\indent

\begin{figure}[thb]
\caption{\protect The total cross section for 
scattering from the ground states of $^{12}$C (a) and $^{208}$Pb (b) to 
(fictitious) $0^+$ states at 15 MeV in $^{12}$N and 
$^{208}$Bi, as a function of outgoing kinetic energy and with transition 
densities proportional to $\delta(r-R)$.  The solid lines are the full DWBA 
results, the dotted lines neglect the Coulomb force completely, and the dashed 
lines are the EMA results.}
\vspace{3mm}
\centering
\leavevmode
\epsfxsize=10cm
\epsfbox{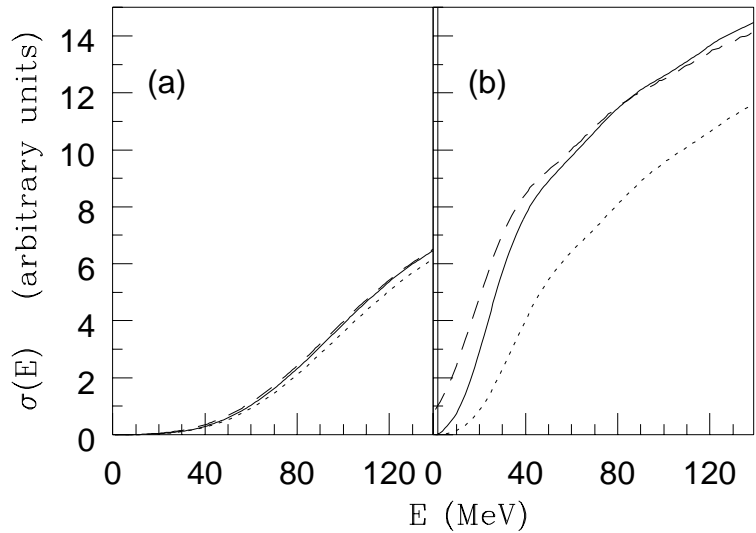}
\label{f:emae}
\end{figure}

At higher energies, the Fermi function doesn't work even for electrons and we
need a different approximation.  In electron scattering, the EMA is used to
untangle the electrostatic attraction of the electron to the
nucleus from the single hard-photon exchange.  As
already mentioned, in its simplest form this approximation consists of
shortening the electron wavelength $k$ to $k_{\rm eff}$ inside the nucleus,
resulting in a larger effective momentum transfer, and rescaling the amplitude
of the wave function by $k_{\rm eff}/k$.  The procedure's accuracy
depends on scattering angle and decreases significantly as $Z$ grows; it is
not regarded as good in lead.  But the application of the EMA to electrons
produced by neutrino scattering, which is straightforward, should work better
than its application in electron scattering.  One reason is that only the 
outgoing wave
experiences a Coulomb force; a more significant one is that the range of the
weak interaction is short.  In electron scattering regions far from the
nucleus, where the local momentum is quite different from $k_{\rm eff}/k$,
affect the scattering matrix element because of long-range photon
exchange.  Here by contrast the only effects are from inside the nucleus.
Figure \ref{f:emae} shows the total cross section for scattering from the
ground states of ${}^{12}$C and $^{208}$Pb to Goldhaber-Teller (analog)
resonances in ${}^{12}$N and $^{208}$Bi, along with the EMA predictions.  The
approximation is quite good above about 30 MeV in lead, leaving only a small
kinematic window in which neither the EMA nor a Fermi function works.  For
electrons, therefore, one can accurately account for the effects of the
electrostatic potential without using the DWBA except in restricted kinematic
regions in heavy nuclei.

Muons are another story however; the best way to extend the EMA to massive
particles is not immediately clear.  The approximation and corrections to it
have been derived from the high-energy limit of the Dirac equation, either
through an eikonal approximation\cite{r:Yennie} or an expansion in inverse
powers of $k$\cite{r:Lenz}.  Both approaches begin by neglecting the mass of
the electron and so separating the usually coupled left and right handed
spinors, hence simplifying the equation.  Here we are concerned with muon
energies below a GeV, for which the rest mass is not negligible.  The basic
ideas behind the simplest approximation --- that the Coulomb potential
shortens the wavelength in the interaction region and increases the wave
amplitude by focusing or defocusing the particles as they approach or leave
--- do not appear to hinge on the particles' rest mass, as long as the
wavelength is short enough for a local momentum to have some meaning.  Some
version of the EMA should therefore apply to intermediate-energy muons as well
as electrons.

The shortening of the wavelength will clearly work the same way for the two
kinds of particle, the only difference being the kinematic relation between
$k_{\rm eff}$ and $E_{\rm eff}$, but what about the amplitude rescaling?
Should it still be $k_{\rm eff}/k$ or some other factor that involves the muon
mass?  We can decide without extending the the full analysis of Refs.\
\cite{r:Yennie,r:Lenz} by viewing the change in amplitude as follows:  The use 
of a
plane-wave approximation in the interaction region is
equivalent to the assumption that the Coulomb potential does not cause the
particles to alter their direction very much when they approach or leave the
nucleus.  It should therefore not strongly alter an outgoing muon wave packet,
which asymptotically is spherical, after it leaves the nucleus except by
slowing it down and thereby changing the average radial wavelength and 
amplitude as the wave moves to larger $r$. (In particular, it should not
cause much reflection.)  But the differential cross section is directly 
related to the radial integral of the square of the wave function over all $r$ 
at a fixed solid angle (in the large time limit), and under the assumptions 
above the Coulomb interaction should
not affect this quantity much once the outgoing wave moves beyond the
nucleus.  Since most of the effect it does have will be to redistribute flux
from one angle to another, the total cross section should be nearly the same
as if the wave kept the same radial wavelength it had inside the
nucleus, i.e.\ as if the potential were equal to $V(0)$ everywhere in space.
For that situation the plane-wave approximation is exact to lowest order in
$Z\alpha$ if the effective momentum and energy $k_{\rm eff}$ and $E_{\rm eff}$
replace the real quantities $k$ and $E$ everywhere in Eq.\ (\ref{e:1}).  In
other words, besides shortening the wavelength in the matrix element ${\cal
M}$, one should replace the phase space factor $k E$ by $k_{\rm eff} E_{\rm
eff}$\cite{r:Kolbe}.  In the Born approximation this is equivalent to a
change in the muon wave function
\begin{equation} 
e^{i\vec{k} \cdot \vec{r}} \longrightarrow \sqrt{\frac{k_{\rm eff} E_{\rm 
eff}}{kE}} e^{i\vec{k}_{\rm eff} \cdot \vec{r}} ~~, 
\end{equation}
i.e.\ a rescaling of the amplitude by $\sqrt{k_{\rm eff} E_{\rm eff}/kE}$ 
rather 
than $k_{\rm eff}/k$.  I will call this approximation the ``modified EMA".

\begin{figure}[htb]
\caption{\protect Total cross sections for several multipole excitations 
(labeling the 
panels) from the ground state of $^{208}$Pb to fictitious states at 15 MeV in 
$^{208}$Bi, as a function of outgoing kinetic energy and with transition 
densities proportional to $\delta(r-R) Y_{JM}$.  The solid lines are the 
results of full DWBA, the dotted lines those with the Coulomb force neglected, 
the dashed lines those of the usual EMA used in electron scattering, and the 
dot-dashed lines those of the modified EMA.}  
\vspace{3mm}
\centering
\leavevmode
\epsfxsize=15cm
\epsfbox{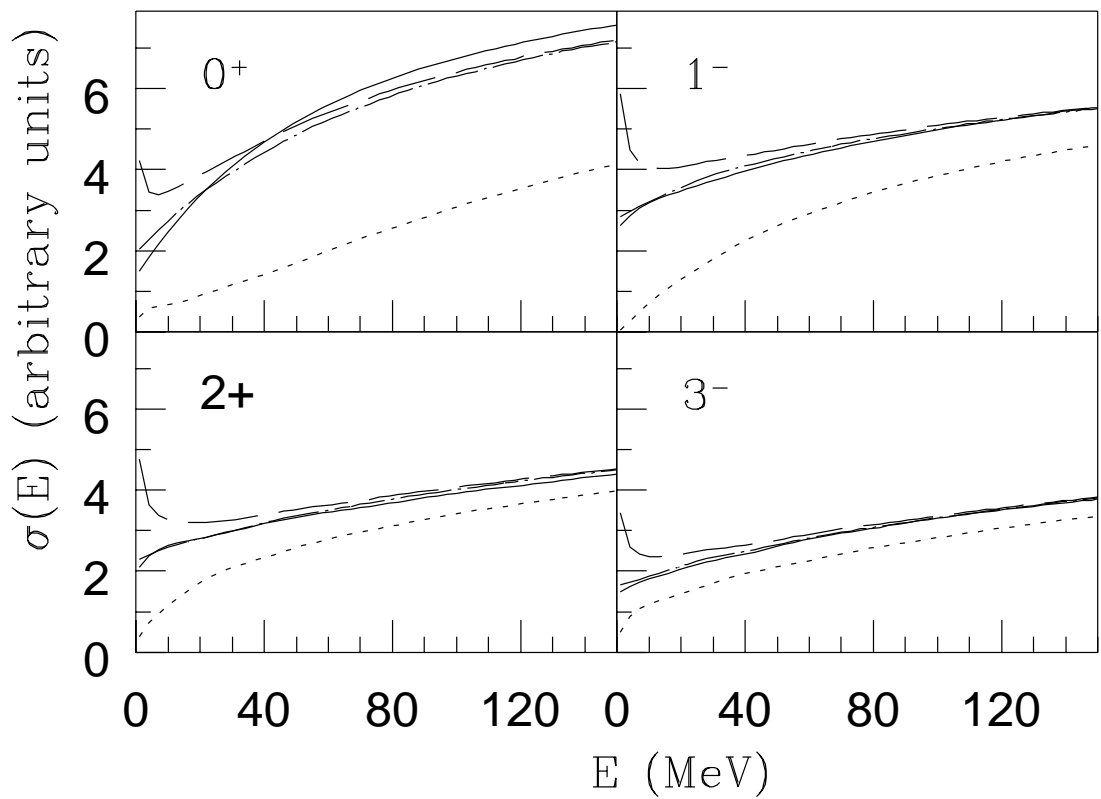}
\label{f:emamu}
\end{figure}

With this alteration the EMA in fact generally works better for muons than for
electrons at low energies for the same reason the Fermi function doesn't work
as well:  the muon mass guarantees that $k_{\rm eff} R$ never drops below
$\sqrt{2 M_{\mu} V(0)} R \approx 0.5$ in nitrogen and 2.6 in bismuth.  Figure
\ref{f:emamu} shows total cross sections for exciting several states with 
different angular momenta in Bi (again
assuming a surface peaked transition density) alongside the modified EMA
predictions and those of the usual EMA, in which the plane wave is scaled by
$k_{\rm eff}/k$.  The usual EMA is pathological at low energies, but the
modified EMA agrees extremely well with the exact solutions except for the
$0^+$ multipole, where it is off by a little over 5\% at muon energies
of 150 MeV.  (There are fewer ways of coupling neutrino and muon angular
momenta to $0^+$, which may accentuate errors in any one partial wave for that
multipole.)  The differences in lighter nuclei are less; in $^{56}$Fe the
agreement is almost exact for all multipoles except at very low energies,
where the $0^+$ multipole still is somewhat worse than the others.  One might
see similar behavior in the $1^+$ multipole, also ``allowed", were the entire
current included.  But the modified EMA is still dramatically better
than the usual plane-wave treatment, even for $0^+$ transitions in heavy
nuclei.

Thus far the only significant laboratory experiments with muon neutrinos in
this energy range are the LAMPF measurements of the cross section for
neutrinos produced by the decay of pions in flight scattering from $^{12}$C.
Because $Z$ is small, electrostatic effects are likewise small.  Most of the
difference between the DWBA and plane-wave approximations in this light
nucleus was correctly accounted for in Ref.\ \cite{r:Kolbe1}, which rescaled
the muon phase space as prescribed by the modified EMA without altering the
nuclear matrix element (though the definition of $k_{\rm eff}$ is reported
incorrectly in the manuscript).  Another situation worth examining, however,
is the scattering of atmospheric muon and electron neutrinos from the iron in
the SOUDAN proton-decay detector.  Results of experiments there and at other
places imply that the ratio of muon-neutrino flux to electron-neutrino flux in
the atmosphere is about 50-80\% of the expected ratio\cite{r:atmonu2}.  The
analyses use the Fermi-gas model of the nucleus and don't generally account
for final-state effects in quasielastic scattering.  Table 1 shows the changes
due to the electrostatic force (calculated in the modified EMA) from the
simple Fermi-gas model predictions for the quasielastic production of muons,
antimuons, electrons, and positrons in iron.  I have chosen relatively low
outgoing momenta and taken neutrino fluxes from Ref.\ \cite{r:flux}.  The
individual event rates actually change significantly for these momenta, but as
is usually the case, the ratio of total $\mu$ to total $e$ events hardly
budges, so that one one apparently can safely neglect Coulomb effects even in
a material as heavy as iron.

\begin{table}
\caption{Calculated event rates for atmospheric neutrinos in $^{54}$Fe, in 
arbitrary units, with and without Coulomb distortion of the final-state 
leptons.
\label{t:1}}
\begin{tabular}{lccc}
\hline\hline
& 100 MeV & 200 MeV & 300 MeV\\
\hline
$\nu_{\mu^-}$ &231 &506 &467 \\
$\nu_{\mu^-}$ (no Coul.) & 165&453 &446\\
$\nu_{e^-}$ &134 &294 &258 \\
$\nu_{e^-}$ (no Coul.) &104 &265 & 247\\
$\nu_{\mu^+}$ &28 &64 &69 \\
$\nu_{\mu^+}$ (no Coul.) & 40&70 &72\\
$\nu_{e^+}$ &24 &38 &38 \\
$\nu_{e^+}$ (no Coul.) &29 &41 &39 \\
$(\nu_{\mu^-}+\nu_{\mu^+})/(\nu_{e^-}+\nu_{e^+})$ &1.65 & 1.72&1.81 \\
$(\nu_{\mu^-}+\nu_{\mu^+})/(\nu_{e^-}+\nu_{e^+})$ (no Coul.)& 1.54&1.70 & 
1.81\\
\hline
\end{tabular}
\end{table}

\section{Conclusions}
\indent

The chief result of this paper is that the effective-momentum approximation
works better in neutrino scattering than in electron scattering, particularly
for muons, where the approach has to be modified slightly to take into account
the charged particle's mass.  It probably doesn't work quite as well for
differential cross sections but preliminary analysis (on $0^+$ transitions
with the vector charge) indicate that the modified EMA is still surprisingly
good.  The real scattering redistributes some flux from forward angles to
places where the cross section is lower, but not as much as one might imagine.
In any event, we are not likely to pay attention to details of
weak-interaction differential cross sections in the near future.  For total
cross sections the modified EMA works admirably for muons down to low energies 
and the usual EMA just as well for electrons except in a region where a Fermi 
function is often adequate.  In heavy nuclei there is a kinematic window for 
electrons in which neither approximation is completely sufficient, but it is 
small. Important examples from this window --- neutrinos from muon decay on 
iodine is one --- are not easy to find at present.

This work was supported by the U.S.\ Department of Energy under
grant DE-FG02-97ER41019.  I thank E. Kolbe and P. Vogel for useful 
discussions and correspondence.

\newpage

\setlength{\baselineskip}{13pt}


\begin{thebibliography}{99}
\bibitem{r:Kolbe} E.\ Kolbe has made this replacement, though without 
shortening the wavelength, in several papers.
\end{thebibliography}
\end{document}